\documentclass[12pt,letterpaper,onecolumn]{article}
\usepackage[latin1]{inputenc}
\usepackage{amsmath}
\usepackage{amsfonts}
\usepackage{amssymb}
\usepackage{braket}
\usepackage[left=1.00in, right=1.00in, top=1.00in, bottom=1.00in]{geometry}

\author{\\
\\
Horace P. Yuen\\
Department of Electrical Engineering and Computer Science\\
Department of Physics and Astronomy\\
Northwestern University, Evanston Il. 60208\\
yuen@eecs.northwestern.edu
}
\title{Fundamental Security Issues in Continuous Variable Quantum Key Distribution}

\begin{document}
\linespread{1}
\maketitle
\linespread{2}
\begin{abstract}
\textit{Several fundamental issues in establishing security in continuous variable quantum key distribution are discussed, in particular on reverse reconciliation and security under heterodyne attack. It appears difficult to derive quantum advantage in a concrete realistic protocol due to source and loss uncertainties, apart from the problem of bounding Eve's information after reconciliation. The necessity of proving robust security for QKD protocols is indicated.}
\end{abstract}
\newpage
\section{INTRODUCTION}

Continuous variable quantum key distribution (CV-QKD) was introduced [1-4] as an alternative QKD approach to the discrete signal BB84 protocol for quantum key generation [5]. The original 3 dB loss limit [1] under beamsplitter attack is extended to arbitrary loss under reverse reconciliation (RR) in [2-4]. In this paper we will point out several fundamental security issues in a finite realistic implementation which are intrinsic to the physics of CV-QKD, as well as the basic cryptographic issue of bounding Eve's information after reconciliation before privacy amplification. It appears that a net key can be generated in CV-QKD not from quantum advantage, but only by classical postdetection processing [5,6], which is very inefficient and in fact its general security has never been fully established. However, coherent detection used in CV-QKD has the advantage of being not susceptible to the detector blinding attacks [7] on single-photon avalanche detectors widely employed in BB84 protocols. Perhaps the combination of coherent detection and BB84 signals in [8-9] would be useful for such purpose. More significantly, CV-QKD brings out clearly the important point that proof of robust security is needed in physical cryptography.

\section{CV-QKD AND RR}

For simplicity we consider the use of coherent states in CV-QKD homodyne protocols. Squeezed states is highly susceptible to all the linear loss in the system and also would, as are other CV-QKD protocols, be subject to all the security points we will make in this paper. Let $T$ be the transmittance of the transmission link which does not include the loss in other system components. Generally our analysis is carried out under the assumption that everything else is perfect unless indicated otherwise.

In CV-QKD the coherent state complex amplitude $x + ip$ is drawn from a circularly symmetric Gaussian distribution of zero mean and variance $V$ in units of 1/4 photon for the field mode, which is the level of noise when $x$ or $p$ is measured in homodyning. The user A sends a long sequence of such pulses, each independently drawn, to user B who checks the noise variance of a randomly chosen subsequence by asking A for the corresponding $x$ or $p$ in each such pulse. This is to establish an advantage, over a possible attacker Eve, on the rest of the pulses to obtain key generation. Then B measures either $x$ or $p$ randomly on the other pulses. Common key bit values are to be derived from reconciliation between A and B's quadrature values before privacy amplification to bring down Eve's information per bit. Mutual information is used to measure the users' advantage over Eve, which has numerous problems in crypto-security, especially for a finite system. See [5,10,11]. However, that is not our real concern here, and the difficulties we will point to cannot be overcome by a mutual information argument.

In \textit{direct reconciliation}, the users derive key bits by B trying to estimate what A sent in the measured quadrature. This is often accomplished by "sliced error correction" [3-4]. There is then a 3 dB loss limit on getting quantum advantage [1], due to Eve's beamsplitter attack in which she splits off half of the signal and delivers the other half via a lossless link to B. Eve would then get an identical copy of the signal at B, and any resulting advantage that may result is of a classical nature from the different noises that B and Eve suffer in their measurements [6,12].

In \textit{reverse reconciliation} (RR), it is A who tries to guess what B has measured. The rationale is that A clearly knows the sent values of $x$ or $p$ better than Eve, and thus an advantage over Eve is assured whe they both try to estimate what B has measured. The sliced error correction protocol for reconciliation is also often used [3-4], though other error correction approaches have been suggested [5]. Indeed, it is often maintained that RR would "in principle" ensure net key generation for any value of $T$.

To represent the RR situation, let $m$ be the parameter value of $x$ or $p$ that A intends for the pulse. Let $m_B$ and $m_E$ be the homodyne result of B and Eve's measurement at the receiving and split-off part of the signal. Then
\begin{equation}
m_B = \sqrt{T}m + n_B
\end{equation}
\begin{equation}
m_E = \sqrt{1 - T}m + n_E
\end{equation}
where the additive noises  $n_B$ and $n_E$ have
\begin{equation}
var \; n_B = var \; n_E = 1
\end{equation}
It is clear the users have no advantage in direct reconciliation under the beamsplitter attack (1)-(2) when $T \leq \frac{1}{2}$.

Even if the parameters $T$ and $m$ are precisely known to the users, it is clear from (1)-(3) that for a sufficiently small $T$ the user A would just be guessing at the value of $n_B$ in RR which has \textit{nothing} to do with the signal or any quantum advantage over E! Also, for $T \ll 1$, Eve knows $m$ basically as well as A does. Since the numerical value of an analog quantity does not have physical meaning after a certain number of decimal places, and indeed is often known to only, say within 1\% of its nominal value, one may say instead there is very much an in-principle limit on $T$ in RR also.

Equally significantly, if a system does not possess robustness in performance within the parameter accuracy limit, it is unstable to operate and is not a viable design for real world applications. In the case of QKD, there is the added issue of \textit{false alarm} that has never been quantitatively studied, in which the users abort a round although there is no Eve present. In the case of a "supersensitive" (very much not robust) protocol, false alarm would occur so frequently that it would prevent its operation altogether.

In the following we will consider the situation in which Eve launches a heterodyne intercept-resend attack near the transmitter, which does not require the unrealistic lossless link replacement underlying (1)-(2) and is readily implementable. We will also indicate that the leak of information to Eve from reconciliation in QKD has never been correctly quantified [11]. Indeed, from (1)-(3) it appears that even for moderate values of $T$, in RR Eve would learn about $m_B$ as much as A  would up to many decimal places, in either open exchange or forward error correction. Certainly there is no justification that the leak is given by the usual $leak_{EC}$ expression [4] which is not generally applicable to begin with.

\section{HETERODYNE ATTACK AND ROBUSTNESS}

Eve can easily launch an intercept-resend attack by heterodyne measurement on both x and p near the transmitter and resend the measured value as the complex amplitude in a coherent state through the link. Again consider RR with $m$ being A's launching value of B's chosen quadrature. As explained in the following, in a realistic system A only knows $m$ as $m_A$ to within an additive noise $n_A$, the variance of which depends on exactly what A does at the transmitter. In place of (1)-(2) we now have, with $var \; \tilde{n}_{E} = 2$,
\begin{equation}
m_B = \sqrt{T}(m + \tilde{n}_E) + n_B
\end{equation}
\begin{equation}
m_E = m + \tilde{n}_E
\end{equation}
\begin{equation}
m_A = m + n_A
\end{equation}
It is clear from (4)-(6) that Eve could determine $m$ better than B in direct reconciliation. In RR, Eve could determine $m_B$ better than A from whatever reconciliation algorithm even when $n_A = 0$. Thus security is totally breached unless Eve is caught heterodyning. However, the latter seems practically impossible and the cryptosystem is in principle supersensitive, because link and other system component losses need to be very accurately determined and, moreover, to remain the same over all the pulses in the round, as follows.

First, consider the $V$ value chosen by A that leads to $m$ in a coherent state from a laser, with $m^2$ far lower than the laser output intensity when it is operating far above threshold for suppressed intensity fluctuation. Large $m$ values would mean the added unity heterodyne noise compared to homodyne is totally insignificant. Large laser output needs to be strongly attenuated to get the chosen $V$, the variance of $m$. For sufficiently small attenuation coefficient the relative fluctuation of its value is huge and A cannot know $m$ at all to within a few units of quantum noise. It appears necessary to split the pulse and homodyne on one part to determine the quadrature value on the transmitted pulse. In fact, due to the relatively large (typically a few percent) margin of accuracy in all laser output power, though it is stable over a long duration, such measurement by A is necessary for pinning down $m$. Thus $n_A$ is at least one unit strong even if there is no other imperfection such as the value of the beamsplitter transmittance.

The \textit{serious} problem, however, comes from the accuracy of loss and $m$ values. To check the presence of Eve's heterodyne attack, B would use (4) and (6) to detect possible added fluctuation from $\sqrt{T} \tilde{n}_E$. It has been appreciated from the beginning of CV-QKD that excess noise and homodyne detection efficiency need to be carefully monitored [5,13]. An experiment on heterodyne attack was carried out [14]. What seems not appreciated is that all the loss values need to be very accurate and stable. To include other system losses such as detector quantum efficiency and homodyne efficiency, one can interprete $T$ in (4) as the total transmittance instead of just link loss. If $T$ deviates from the presumed value by 2\%, an $m^2 = 25$ photons commonly employed [3] would imply Eve's heterodyne attack cannot be detected even apart from finite statistical fluctuation. For a proper security analysis, one would need to represent exactly how each loss parameter enters into the system representation, with finite fluctuation fully accounted for in a complete statistical analysis. For example, if A uses a 50/50 beamsplitter to determine the $m$ value after the laser output is strongly attenuated, the beamsplitter transmittance cannot be smaller by 0.1\% for $m = 10$ and total 10 dB user loss in order not to confuse Eve's added noise as part of the signal. In the other direction, \textit{false alarm} would result. It does not appear such a supersensitive cryptosystem can be practically similar to the cases of classical resolution beyond the diffraction limit and singular detection in non-white Gaussian noise.

\section{OTHER SERIOUS ISSUES}

A most serious problem is Eve's information gain from reconciliation. Under (1)-(2) in which Eve's presence cannot be detected in principle, Eve still would learn a lot since $m_E$ and $m_A$ from (2) and (6) imply Eve and A are in similar positions with respect to $m_B$, that there is really no advantage to A. Even when $n_A = 0$, the advantage seems small and it is not clear why the users can derive a net key in a finite protocol. The problem can be traced in part to the lack of a quantitative bound on Eve's information gain in any reconciliation procedure [11], except asymptotically for linear error correcting codes under collective attacks in BB84. There is no definite proven result for CV-QKD either in [4].

Collective attack, however, is very unreasonably restrictive in both BB84 and CV-QKD. For example, when Eve attacks a portion of the pulses and leaves the rest intact, it is outside the scope of collective attack. On the other hand, Eve may well choose to do that to escape detection while gaining a considerable amount of information. It is intuitively impossible that collective attack is optimal for Eve, and the known "proofs" that it is are not valid. This and the many general inadequacies of QKD security proofs in BB84 and CV-QKD are to be treated elsewhere, while a glimpse can be found in [11].

\section{ROBUST SECURITY}
Robustness with respect to parameter accuracy and fluctuation is crucial in engineering systems. Quantum information system is often very sensitive to disturbance, which is likely one main reason why it is so difficult to realize experimentally. Security is something that can only be established theoretically if at all, because there
are an unlimited number of possible attack scenarios and the history of cryptography shows that relying on security from past experience can be dangerous. In physical cryptography such as QKD where security depends on physics at least as much as on mathematics, it is therefore mandatory to establish robust security carefully on a \textit{complete} representation of the physical cryptosystem. This has not been done in CV-QKD or BB84, and must be addressed if QKD is ever going to be practically useful.

\section*{ACKNOWLEDGEMENT}
This work was supported by the Air Force Office of Scientific Research. I would like to thank T. Hirano for useful discussions.


\begin{thebibliography}{9}
\bibitem{ref1} F. Grosshans and P. Grangier, Phys. Rev. Lett. 88, 057902 (2002)
\bibitem{ref2} F. Grosshans and P. Grangier, arXiv:0204127, 2002.
\bibitem{ref3} F. Grosshans, G. Van Assche, J. Wngler, R. Tualle-Brouri, N. Cerf, and P. Grangier, Nature 421, 238, 2003.
\bibitem{ref4} G. Van Assche, J. Cardinal, and N. Cerf, IEEE Trans. Inform. Theory 50, 394, 2004.
\bibitem{ref5} V. Scarani, H. Bechmann-Pasquinucci, N. Cerf, M. Dusek, N. Lutkenhaus, and M. Peev, Rev. Mod. Phys. 81, 1301, 2009.
\bibitem{ref6} H. P. Yuen and A. Kim, Phys. Lett. A 241, 135, 1998.
\bibitem{ref7} I. Gerhardt, Q. Liu, A. Lamas-Linares, J. Skaar, C. Kurtseifer, and V. Makarov, Nat. Commun. 2, 349, 2011.
\bibitem{ref8} H. P. Yuen, in Proceedings of the Fourth International Conference on Squeezed States and Uncertainty Relations, NASA Conference Publication 3322, pp. 363-368, 1995.
\bibitem{ref9} H. P. Yuen, Quantum Communication, Computing, and Measurement 2, ed. by Kumar et al, Luwer Academic/Plenum Publisher, New York, pp. 399-404, 2000.
\bibitem{ref10} H. P. Yuen, IEEE J. Sel. Topics in Quant. Electron. 15, 1630, 2009.
\bibitem{ref11} H. P. Yuen, arXiv:1205.3820, 2012.
\bibitem{ref12} U. M. Maurer, IEEE Trans. Inform. Theory 39, 733, 1993.
\bibitem{ref13} R. Namiki and T. Hirano, Phys. Rev. Lett. 92, 117901 (2002).
\bibitem{ref14} J. Lodewyck, T. Debuisschert, R. Garcia-Patron, R. Tualle-Brouri, N. Cerf, and P. Grangier, Phys. Rev. Lett. 98, 030503 (2007).

\end{thebibliography}
\end{document}